\documentclass[preprint,12pt,authoryear]{elsarticle}

\usepackage{amssymb}
\usepackage{amsthm}

\theoremstyle{definition}
\newtheorem{definition}{Definition}[section]

\usepackage{graphicx}
\usepackage{subfig}
\usepackage{multirow}
\usepackage{eurosym}

\usepackage[
    type={CC},
    modifier={by-nc-nd},
    version={4.0},
]{doclicense}

\journal{Journal of Systems and Software}

\begin{document}
\doclicenseThis
\begin{frontmatter}


\title{A complex network analysis of the Comprehensive R Archive Network (CRAN) package ecosystem}

\author{Mar\c{c}al Mora-Cantallops\corref{cor1}\fnref{label2}}
\cortext[cor1]{Corresponding author}
\ead{marcal.mora@uah.es}
\author{Salvador S\'anchez-Alonso\fnref{label2}}
\author{Elena Garc\'ia-Barriocanal\fnref{label2}}

\fntext[label2]{\{marcal.mora, salvador.sanchez, elena.garciab\}@uah.es}
\address{Universidad de Alcal\'a, Spain}


\author{}

\address{}

\begin{abstract}
Free and open source software package ecosystems have existed for a long time and are among the most sophisticated human-made systems. One of the oldest and most popular software package ecosystems is CRAN, the repository of packages of the statistical language R, which is also one of the most popular environments for statistical computing nowadays. CRAN stores a large number of packages that are updated regularly and depend on a number of other packages in a complex graph of relations; such graph is empirically studied from the perspective of complex network analysis (CNA) in the current article, showing how network theory and measures proposed by previous work can help profiling the ecosystem and detecting strengths, good practices and potential risks in three perspectives: macroscopic properties of the ecosystem (structure and complexity of the network), microscopic properties of individual packages (represented as nodes), and modular properties (community detection). Results show how complex network analysis tools can be used to assess a package ecosystem and, in particular, that of CRAN.
\end{abstract}




\end{frontmatter}


\section{Introduction}
\label{intro}

The surge in open source software (OSS) development has resulted in abundant available software packages that, in each particular software ecosystem, can be used by developers as building blocks for new projects, reducing development costs and time~\citep{mohagheghi} and which can contribute with a positive and significant value-added return~\citep{nagle2019open}. In a recent report, the European Commission report estimated that using free/libre and open source software (FLOSS) saves the European economy roughly \euro114 billion per year directly and up to \euro399 billion per year overall~\citep{harutyunyan}. But, on the other hand, such third-party libraries introduce both direct dependencies and transitive dependencies that need to be kept updated to prevent vulnerabilities and bug propagation that might endanger the whole ecosystem~\citep{cox}. Although developers can have a clear vision of the direct dependencies they add to their packages, transitive dependencies might be less clear as they are not included by them, becoming hidden one or multiple levels below the direct dependency. Even the common action of updating packages entails risks, as changes might break existing functionalities on other packages~\citep{raemaekers}.

One of the oldest and most popular software package ecosystems is CRAN, the repository of packages of the statistical language R. The R programming language is widely used among statisticians and data miners for developing statistical and data analysis libraries, while also being one of the most popular languages among data scientists thanks to its flexibility and expansion capabilities, as R can be
extended through user-created packages. As of March 2020, it ranks 11th in the TIOBE index (https://www.tiobe.com/tiobe-index/), a measure of popularity of
general purpose programming languages. The Comprehensive R Archive Network (CRAN) (https://cran.r-project.org/) is a network of web servers around the world where R source code, R manuals, documentation, and contributed packages can be found, and it can be considered as the official repository, containing the largest collection of available R packages. At the end of 2019, it hosted a total of 15.368 packages.

As is common in these environments, developers of many software applications or packages rely on using other OSS packages; such dependencies manifest in different forms. In some cases, packages or applications might need the source code of another package or class to compile correctly. In other cases, such as CRAN, source-code dependencies do not exist; binary-level library sharing is required for many package to function properly. Such dependencies might be shared among many projects and repositories, although in CRAN this is limited to its own repository except for a few and specific packages that are stored in Bioconductor. A package management system serves the purpose of managing such dependencies, which is important for both functioning and maintenance (e.g., automated updating) of software packages. \citet{german} conducted an exploratory empirical study on the evolution of the R software ecosystem, and showed how R was "a flourishing ecosystem of user-contributed packages" that was growing and contained a "strong set of core packages". Among their observations, they found packages to be typically well-maintained.

Open-source software ecosystems such as CRAN could be considered as very complex networks of artifacts, due to the increase in collaborative development under the open source software paradigm in the last two decades. This enables us to study software package ecosystems from the perspective of complex network analysis (CNA) to analyse its structural, individual and modular characteristics, but also to detect potential risks and vulnerabilities in the network formed by such packages. 

Overall, the current analysis aims to demonstrate how complex network analysis techniques can be applied to a OSS package ecosystems (such as CRAN) after building its dependency network, and how the results reflect its scale-free and small-world behaviour, the potential vulnerability of some of its packages and the modular structure that is hidden behind the dependency network. 

The remainder of this article is structured as follows. Section 2 discusses related work. Section 3 includes details on the data source and the network construction. Section 4 addresses the results of the complex network analysis on CRAN. Section 5 adds perspective to the results, linking them to the related work and discussing this work's contributions. Section 6 concludes, with an additional mention to threats to validity and future lines of work in section 7.

\section{Background}
\label{background}

The networks formed by real-world systems in many disciplines (e.g. sociology, biology or computer science, among many others) have been proven to exhibit complex network properties, such as being scale-free~\citep{barabasi}, following a power law distribution and obeying the small world principles~\citep{watts}. In the past decade, the growth in the body of the literature studying complex networks is remarkable in a wide variety of fields. Some examples include politics~\citep{halberstam2016, mora2019}, medicine~\citep{rubinov2010}, economics~\citep{cerina2015}, trade networks~\citep{kasakawa2016}, transport~\citep{feng2017}, or even online gaming~\citep{mora2018}. Software systems, represent another important subset of systems that can benefit from complex network analysis~\citep{myers2003}.

Earlier work has revealed that class dependency graphs of individual open source software systems exhibit complex network characteristics, not only in their scale-free degree distributions and the existence of small-world phenomena, but also in their community structure, as ~\citet{vsubelj2011} emipirically confirmed in their study of several networks constructed from Java and various third party libraries. To do so, they built class dependency networks, where nodes represent software classes and edges represent dependencies among them.~\citet{pan2011} used complex network theory as a tool to analyze the evolution of object-oriented software from a multi-granularity perspective. Besides other relevant findings, they also highlighted how complex network techniques provided ``a different dimension to our understanding of software evolution and also are very useful for the design and development of object-oriented software systems". In the same line of work, \citet{chong} enhanced this technique by assigning weights to the edges of the network ``to denote the strength of communicational cohesion between a pair of related software components", all in order to capture its structural characteristics and to enable a maintainability and reliability analysis. They found that this technique made pattern identification easier and that it was also possible to identify software components that violated common software design principles. This idea of using complex network analysis (via dependency graphs) was also applied by~\citet{zimmermann2008} to evaluate Windows Server 2003; their models improved the performance of models build only by complexity metrics (without network metrics) by ten points and were able to identify twice as many critical binaries.~\citet{li2018} recently noted how ``traditional software reliability evaluation approaches lack the analysis of inter-component interactions of component‐based software systems" and proposed a reliability evaluation model for such systems based on complex network analysis.

This software systems approach has been extended to software package networks over the last decade.~\citet{zheng}, for example, recognized how ``understanding the structure of software systems can provide useful insights into software engineering efforts and can potentially help the development of complex system models applicable to other domains". To prove their hypothesis, they empirically analyzed the package ecosystem of the Gentoo Linux distribution, modeling software packages as nodes and their dependencies as nodes, and developed two growth models for the network. In their future work, they stated that to ``study a number of open-source software systems beyond Gentoo Linux [...] could lead to fruitful research contributions". One of such contributions, by \citet{fortuna2011}, compiled all packages and dependencies/conflicts from the Debian/GNU operating system per each major stable release and discussed the parallelisms between its evolution and dynamics over the first 10 releases with that of ecological webs of interacting species, demonstrating the interdisciplinary nature of the CNA toolset. ~\citet{abate2009} introduced novel notions on dependencies and sensitivity (related to how critical a component is); the main applications for these metrics were ``tools for quality assurance in large component ecosystems and upgrade risk evaluation" that they applied to the Debian package ecosystem; among other findings, they found Debian to be ``a small world".

\citet{cataldo}, in the editorial of a topical issue devoted to the complex network perspective on software engineering, stated that network-based methods can be utilized to study research questions relevant to empirical software engineering. ~\citet{zheng} wrote, a few years prior, that the reasons behind the lack of CNA studies on software engineering were ``the difficulties with data collection and the lack of applicable models". \citet{cataldo} recognized that, with more massive data sets from platforms such as SourceForge or gitHub (to which many other repositories could also be added), it would mark ``the beginning of a fruitful field of research".

R and CRAN have also been the objects of research by a few studies from the perspective of their packages.~\citet{decan2016} made use of the dependency network to compute dependencies in their study and comparison of three different ecosystems (R's CRAN archive network, Python's PyPI distribution, and JavaScript's NPM package manager), but did not follow a complex network analysis perspective. In the future work of a later article, however, ~\citet{decan2019} hinted at the how dependency networks of open source packaging ecosystems also reveal complex network behavior and that ``it would be worthwhile to study [...] the complex network properties of ecosystem package dependency network". It is also worth adding, although this perspective will not be used in the current article, that complex (and social) network analysis can also be useful to assess the contributor networks and communities that take part in the ecosystem, as modeled by~\cite{korkmaz2018}. 

This work, thus, attempts to understand whether CNA can be applied to analyse a software ecosystem such as CRAN, pointing at additional analytical tools and opening new possibilities for developers and software engineers in general when assessing package dependency networks, their structure or their quality.

\section{Materials and methods}
\label{methods}

\subsection{Data extraction}
\label{data}

The extraction was executed using R and the ``pkgsearch'' package~\citep{pkgsearch}, which uses the `R-hub' search server (see https://r-pkg.org) and the CRAN metadata database to provide detailed information about CRAN packages. The extracted metadata per package includes the following key elements among others:

\begin{itemize}
\item
  Descriptive features, such as name of the package, description or
  version.
\item
  Author(s).
\item
  Imports: dependencies that are required for the package to work.
\item
  Suggests: packages that can be used by the package but that are not
  required.
\item
  Depends: currently states the version of R required by the package,
  but it is relevant as before R 2.14.0 this field contained the
  dependencies to other packages (therefore, it was equivalent to
  imports).
\item
  Date/Publication: time stamp with the date of publication of the
  particular version of the package.
\end{itemize}

The extraction results in a total of 15.368 unique packages as of the
31st of December 2019. 148 additional external packages come from dependencies that are stored in another related repository, Bioconductor (an open source and open development software project for the analysis and comprehension of genomic data). Thus, a total 15.516 packages are considered. 

Although the extraction using the ``pkgsearch'' package covers most of
the features needed for the following analysis, data was manually
cross-checked to ensure the reliability of the package and the obtained
information was complemented with the information directly scraped from
the CRAN web repository at https://cran.r-project.org. The number
of packages obtained using this method is exactly the same and
the results were positive, so we can depart from the assumption that the
information obtained from ``pkgsearch'' is reliable.

\begin{figure*}
\centering
  \includegraphics[width=\textwidth]{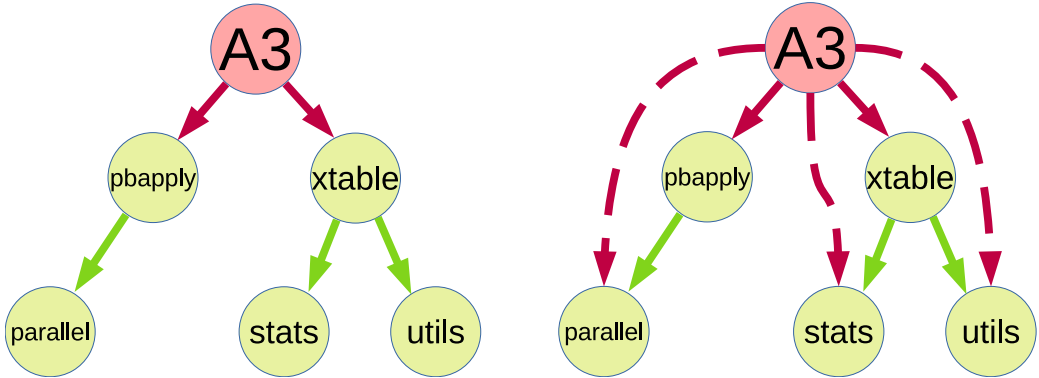}
\caption{Dependency graph of the 'A3' package (left) compared to its transitive closure (right).}
\label{fig:a3}       
\end{figure*}

\begin{table}
\small
\centering
\caption{Properties of the CRAN package dependency network}
\label{tab:1}       
\begin{tabular}{|l|c|c|c|}
\hline
Network & Nodes (N) & Edges (E) & Avg. Degree (k) \\
\hline
Full CRAN (FC) & 15516 & 66594 & 8.584 \\
Giant Component (GC) & 13838 & 66574 & 9.622\\
Transitive Closure (TC) & 13838 & 381998 & 55.210\\
TC except base packages (TCNB) & 12686 & 274449 & 43.26\\
\hline
\end{tabular}
\end{table}

\subsection{Network construction}
\label{network}

Inspired by the same principles as previous works on class dependency networks~\citep{vsubelj2011,vsubelj2012,chong} and following the approach used in similar research on package ecosystems or open-source software systems, the CRAN network will be represented by the packages as nodes and their dependencies as edges~\citep{zheng,kikas}. Note that prior to the rollout of namespaces in R 2.14.0, the metadata field ``Depends" was the only way to reflect dependencies on another package. After that, developers are expected to use the field ``Imports" instead; to both account for older and newer packages, and possible inappropriate labeling in the metadata, both fields are combined to obtain the complete dependencies.

Formally, CRAN can be considered to be a set of packages $P = {P_1, P_2, \cdots}$. The package dependency network is, thus, a directed graph $DG(N,E)$ with nodes $N$ and edges $E$ where node $i$ corresponds to package $P_i$ and directed links $(i,j) \in E$ represent a dependency between packages $P_i$ and $P_j$; $P_i$ imports or depends on $P_j$. The average number of edges directed towards the network nodes is their average in-degree ($k^{in}$), while the average number of edges leaving them become their average out-degree($k^{out}$). The average degree in the network, therefore, can be represented as $k = k^{in} + k^{out}$. Note how $k^{in}_i$ corresponds to the number of classes that use (import or depend on) $P_i$, while $k^{out}_i$ corresponds to the number of other packages that are required for $P_i$ in order to function.

The full CRAN network, constructed using packages as nodes and direct dependencies as edges, is not connected. A number of packages can be found in the periphery of the network that are either standalone (with no dependencies) or depend on a handful of other peripheral packages. Following the approach in previous works, such disconnected packages are discarded by reducing the CRAN network to its largest connected component (also known as the giant component). Additionally, to reflect the transitive dependencies (the recursive dependencies of dependent packages) in the network, the transitive closure of the network is considered. 

\begin{definition}{Transitive closure.}
The transitive closure of $G=(N,E)$ is a graph $G+=(N,E+)$ such that for all $i, j$ in $N$ there is an edge $(i, j)$ in $E+$ if and only if there is a path from $i$ to $j$ in $G$.
\end{definition}

An example of such transformation is represented in figure \ref{fig:a3}, where the ego network of the package 'A3'~\citep{a3} (the first in alphabetical order in CRAN) is used for reference.

Finally, there are two special sets of packages to be considered:
\begin{itemize}
    \item Base packages that are included with the R distribution. This list is formed by a total of 14 packages, namely: `base', `compiler', `datasets', `grDevices', `graphics', `grid', `methods', `parallel', `splines', `stats', `stats4', `tcltk', `tools' and `utils'.
    \item CRAN-recommended add-on packages, included in all binary distributions of R. These are a total of 15 packages: `KernSmooth', `MASS', `Matrix', `boot', `class', `cluster', `codetools', `foreign', `lattice', `mgcv', `nlme', `nnet', `rpart', `spatial' and `survival'.
\end{itemize}

As this set of 29 packages is highly relevant and close to the core (there is no binary distribution of R without them), a disproportional large number of packages depend on them. Although this is indeed relevant to assess the modular structure of the network, for instance, it can distort the analysis of the vulnerabilities. Base packages could hardly be considered third-party risks to the R package ecosystem when they are inseparable from the R base distribution. Therefore, the transitive closure network without these two sets of packages will also be considered for a complete analysis and to obtain insights beyond the core packages of R.

Table \ref{tab:1} shows the properties of the CRAN package dependency network in the four iterations (full network, giant component, transitive closure and transitive closure without base packages) considered above. The latter three will also be compared to their equivalent Erdös-Renyi random graphs~\citep{erdHos}, where a link is placed between a certain pair of nodes with probability $k/(n-1)$, where $k=2e/n$ for a given number of nodes ($n$) and edges ($e$).

\section{Analysis and results}
\label{results}

In this section, the CRAN package dependency network will be analyzed in three dimensions in order to show how complex network analysis can be applied to package ecosystems while benefiting developers, maintainers and contributors. These three dimensions result in the following research questions:
\begin{enumerate}
    \item (RQ1) In regard to its structure and complexity, does the CRAN package dependency network exhibit scale-free and small-world behaviours?
    \item (RQ2) Concerning the individual packages in the ecosystem, what are the most vulnerable?
    \item (RQ3) Is the CRAN package dependency network modular? Is it possible to infer an underlying structure using the relationships between packages?
\end{enumerate}

Each subsequent subsection aims to answer one of these research questions.

\subsection{RQ1: Structure and complexity}
\label{structure}

\begin{table}
\small
\centering
\caption{Network statistics for the CRAN package dependency network.}
\label{tab:2}       
\begin{tabular}{|l|c|c|c|c|c|c|c|}
\hline
Network & $\gamma$ & $C$ & $C_{ER}$ & $l$ & $l_{ER}$ & D & $n_d$\\
\hline
Giant Comp. (GC)& 1.931 & 0.21 & 0.0006 & 3.10 & 4.47 & 0.0003 & 78.9\\
Trans. Closure (TC) & 2.769 & 0.31 & 0.0040 & 2.21 & 2.79 & 0.0020 & 5.6 \\
TC except base (TCNB) & 2.613 & 0.26 & 0.0034 & 2.94 & 2.860 & 0.0017 & 15.5 \\
\hline
\end{tabular}
\end{table}

\begin{figure*}

\begin{minipage}{.5\linewidth}
\centering
\subfloat[]{\label{main:a}\includegraphics[scale=.4]{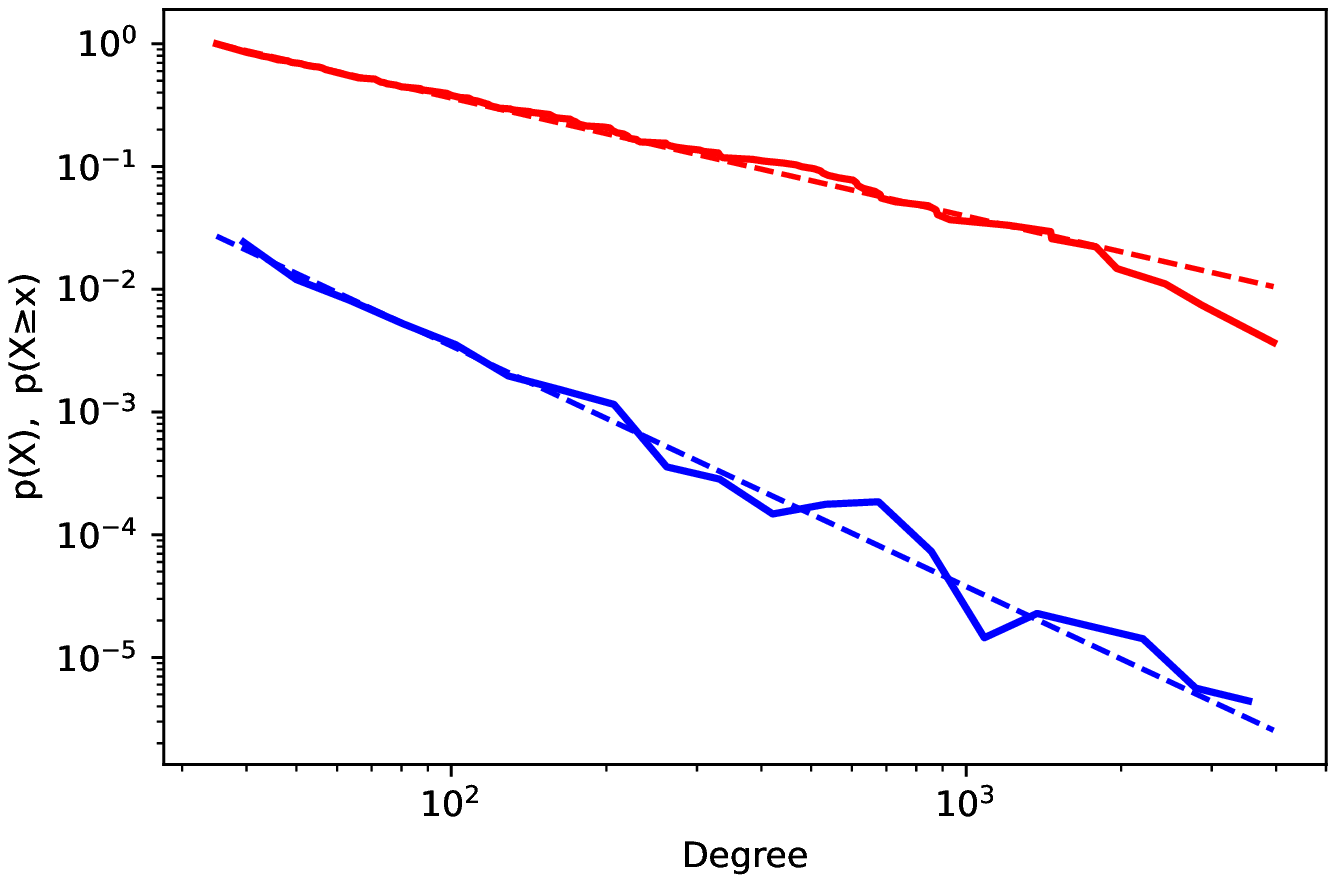}}
\end{minipage}%
\begin{minipage}{.5\linewidth}
\centering
\subfloat[]{\label{main:b}\includegraphics[scale=.4]{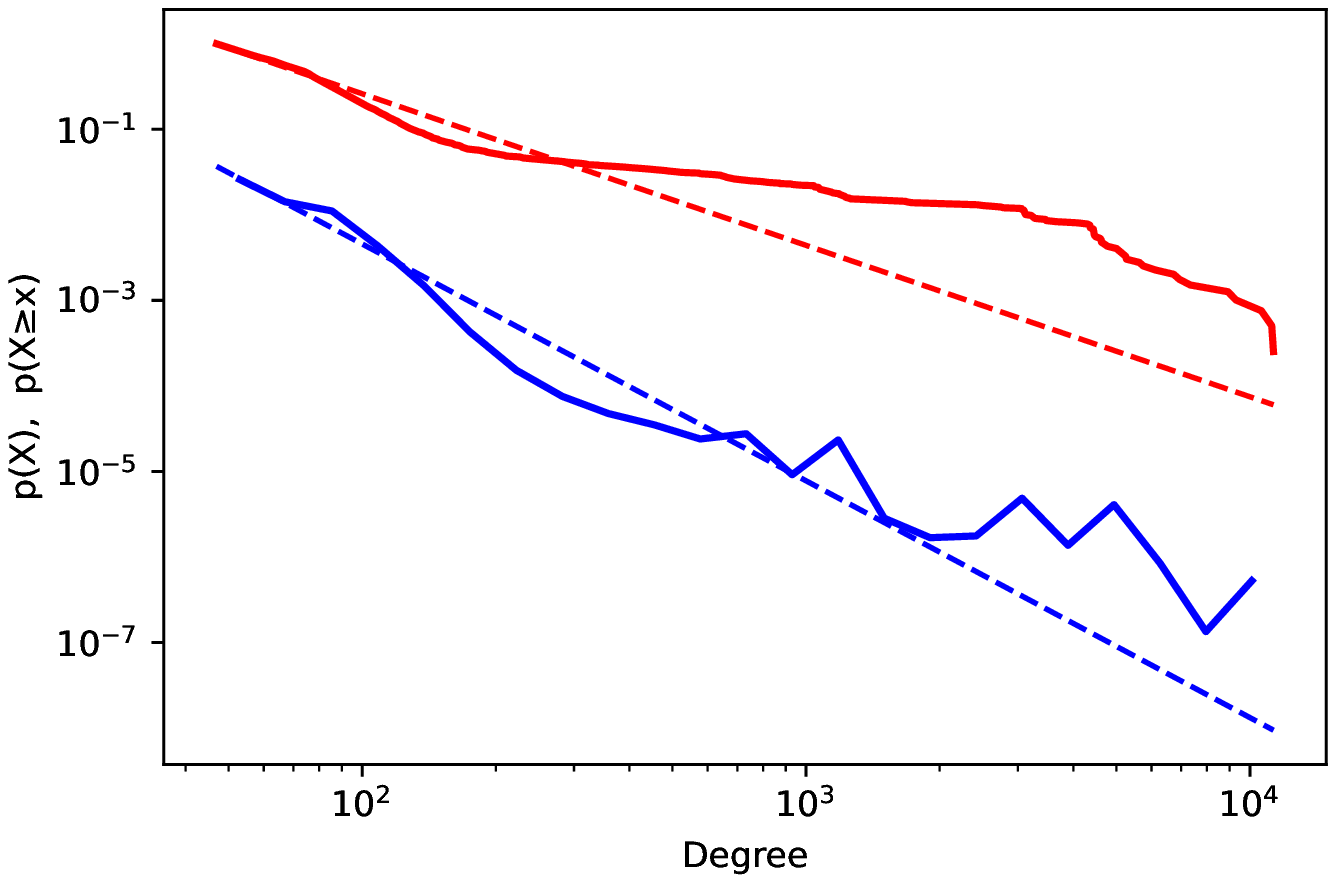}}
\end{minipage}\par\medskip
\centering
\subfloat[]{\label{main:c}\includegraphics[scale=.4]{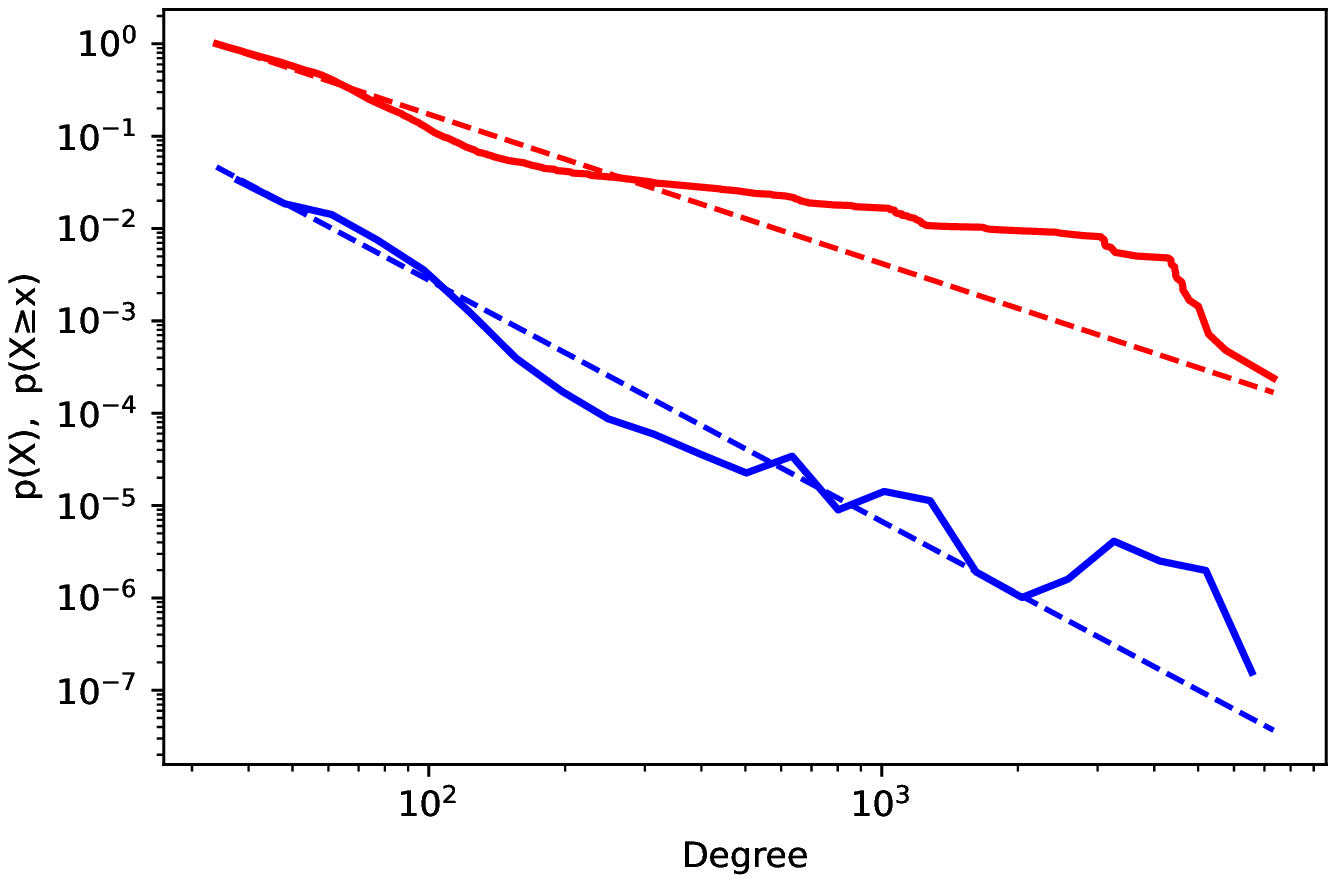}}

\caption{\label{powerlaw}Probability density function ($p(X)$, blue, corresponding to the lower lines in each sub-graph) and complementary cumulative distribution function ($p(X \ge x)$, red, corresponding to the upper lines in each sub-graph) of degrees in (a) Giant Component (GC); (b) Transitive Closure (TC); (c) TC except base packages (TCNB).}
\end{figure*}

Degree distribution experienced by simple random graphs is either binomial or Poisson when the size of the graph is large~\citep{albert2002statistical}. However, many real-world networks have been found to follow different patterns. For instance, many networks’ degree distribution follows the power-law property, while others’ exhibits non-power-law features such as exponential cutoffs~\citep{amaral2000}. Software networks have been found to follow a power-law degree distribution~\citep{potanin2005,vsubelj2012}:

\begin{equation}
p_k \sim k^{-\gamma}
\end{equation}

with $p_k$ as the probability of a certain degree $k$ and $\gamma$ as the scale-free exponent, with $\gamma>1$. The power-law relationship can be directly observed in a log-log plot with an straight line of slope $-\gamma$~\citep{alstott2014powerlaw}. The values for $\gamma$ in each network can be found in table~\ref{tab:2} and their corresponding log-log diagrams in figure~\ref{powerlaw}, where the complementary cumulative distribution function is also added for reference. All three networks exhibit power-law degree distributions, in line with the hypothesis of their scale-free property. In scale-free networks, the probability of two nodes being linked is not a constant as in random graphs; instead, it depends on the number of links that a node already has. In other words, the more popular a node is, the more likely it is to increase its number of links when new nodes are added. In the case of the CRAN ecosystem (and the dependency networks that are being analyzed here), this implies that packages with a high number of reverse dependencies (meaning high in-degree) are more likely to become dependencies of newly developed packages too, a phenomenon that is noticeable, for example, looking at the base packages. 

\begin{figure*}
\centering
  \includegraphics[width=0.75\textwidth]{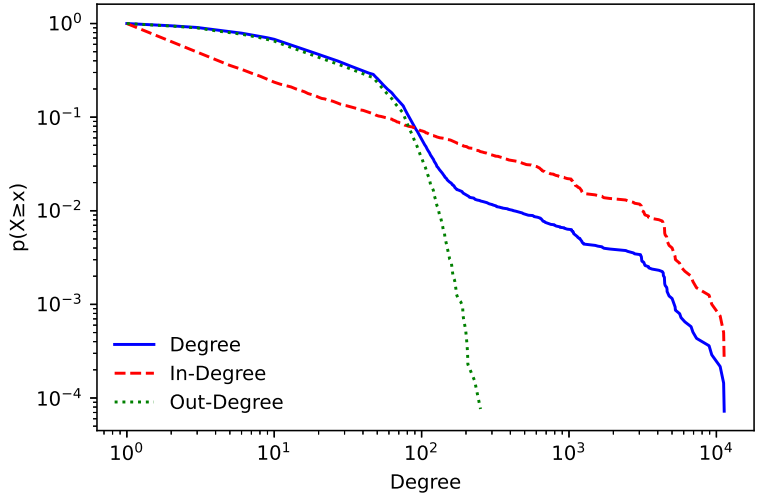}
\caption{Comparison between the degree, in-degree (reverse dependencies) and out-degree (dependencies) distributions. It can be seen how the out-degree distribution (green, dotted) is heavily truncated versus the in-degree (red, dashed) distribution.}
\label{fig:powerlawinout}       
\end{figure*}

Nevertheless, this power-law probability distribution would then be expected only in the in-degree distribution, as it represents the number of other packages that require a given package to function. In-degree distribution, thus, is analogue to the degree of package reusability. Besides, out-degree distribution takes another approach; as it represents the number of packages required for a given package to work, it reflects software complexity. The ideal software project should exhibit scale-free behaviour on in-degree (high reusability) and a highly truncated out-degree distribution (avoiding high complexity in a single package dependencies)~\citep{vsubelj2012}. In the case of CRAN, these distributions are represented in figure \ref{fig:powerlawinout}, showing how both properties are fairly obeyed by its distribution. 

However, while high reusability decreases the probability of fault propagation through the system, it also increases its vulnerability in case of a bug in any of the highly reused packages, as even a very small fraction of faulty nodes can already render the entire system inoperable. Both packages with high in-degree and out-degree need to be particularly monitored and carefully maintained; the details per node will be analyzed in subsection~\ref{nodes}. 

On the other hand, small-world~\citep{watts} behaviour usually refers to high clustering ($C$) and a short average distance ($l$) between the nodes. Clustering measures transitivity in the network; for unweighted graphs, the clustering of a node u is the fraction of possible triangles through that node that exist. It can also be understood as the probability of any two neighbours of a given network being also linked. From their definition, small-world graphs should exhibit $C \gg C_{ER}$ and $l \sim l_{ER}$, where both $C_{ER}$ and $l_{ER}$ are the respective properties for an equivalent Erdös-Renyi random graph. All these figures for the CRAN networks can be found in table~\ref{tab:2} and, from them, it can be derived that the ecosystem also behaves as a small-world network, which is desirable in well designed software projects~\citep{vsubelj2012}, as it indicates a good relationship between packages that share similar functions (measured by $C$) while avoiding a balkanisation of the ecosystem (where parts become independent and unaware of each other) (measured by $l$). It should be noted, though, that these measures only make sense with the dependency graph converted to undirected~\citep{kohring2009complex}; the opposite would actually imply that there would be cyclic dependencies among packages, which are undesired. 

Network densities ($D$), which are the ratio between the actual number of edges and those of a complete graph, were also computed for all three networks in table~\ref{tab:2}; as expected for real-world and software networks~\citep{zheng}, the CRAN network is sparse in general. 

Therefore, in answer to the first research question, the CRAN package dependency network adheres to a power-law, which follows the principles of the scale-free networks, while also reflecting a small-world behaviour.

\subsection{RQ2: Individual packages (nodes)}
\label{nodes}

As developers work with previous packages to develop newer ones, quality, maintenance and trustworthiness of the existing packages is key for the stability of the ecosystem. These properties, however, are not particularly visible and they are not under the developer's control, either. A famous incident happened in 2016 when a single JavaScript package, called \textit{left-pad} (https://github.com/stevemao/left-pad/issues/4) was removed from the central JavaScript package repository \textit{npm}. This removal caused issues not only for the projects that depended on it, but also for those that depended transitively on the package. In the case of CRAN, previous work has shown that up to 41\% of the errors in CRAN packages were caused by incompatible changes in one of its dependencies (direct or transitive)~\citep{claes2014maintainability}. Thus, issues or bugs with packages propagate through any number of levels of dependency, not only on direct ones. It is therefore possible to measure the vulnerability ($v$) of the ecosystem to an issue of a given package as the fraction of packages in the whole ecosystem that would be impacted by the propagation of that issue through its dependencies. Such information could be incorporated in measuring package importance with regards to vulnerability in an ecosystem, as a high vulnerability score should alert developers and maintainers to ensure a fast response to bugs ans issues, as they could both raise a chain reaction and raise the interest of any attacker interested in finding an opportunity to exploit the project~\citep{kikas}.

To find the most critical nodes in regards to vulnerability, one can make use of the centrality metrics, whose main purpose is measuring nodes influence. Many centrality measures are available, and each one defines ``relevancy'' differently. For instance, nodes with high betweenness centrality influence the flow around a system, while closeness centrality aims to measure how well placed a node is in the network. Degree centrality, even though it could be considered as the simplest measure of node connectivity, is also the most appropriate metric to find very connected or popular nodes that, in case of failure, would impact a larger number of other nodes (or packages). In the present case, the normalized degree centrality ($DC_i$) will be computed for each package in the network as

\begin{equation}
DC_i = \frac{k_i}{n-1}
\end{equation}

with $k_i$ being the degree of node $i$, $n$ the total number of nodes in the network and $DC_i \in [0,1]$. We can, thus, assimilate the degree centrality for a node $i$ to the fraction of nodes it is connected to (independently of the direction of the link). For each node, both the degree centrality $DC$ and the corresponding $v$ is represented in table~\ref{tab:3}, combined with their direct dependencies ($DD$) and transitive dependencies ($TD$). The table contains the top 20 influential nodes for both the TC and TCNB cases, as it can be noticed that in the TC case base packages take most of the slots. For the TCNB case, there are a total of 20 packages that, if exposed, would individually impact more than 30\% of the whole network through their dependencies. In this list, it is worth highlighting how a number of packages have quite limited numbers of direct dependencies but their transitive ones are up to three orders of magnitude larger. On the other hand, if the base packages are included, one can realize how issues in `methods', `utils' or `stats' would basically take the whole ecosystem down.

\begin{table}
\small
\centering
\caption{Top 20 influential nodes in CRAN.}
\label{tab:3}       
\begin{tabular}{|c|c|c|c|c|c|c|c|c|c|}
\hline
\multicolumn{5}{|c|}{\textbf{All packages (TC)}} & \multicolumn{5}{|c|}{\textbf{Excluding base packages (TCNB)}} \\
\hline
\textbf{Package} & \textbf{$DD$} & \textbf{$TD$} & \textbf{$DC_i$} & $v$(\%) & \textbf{Package} & \textbf{$DD$} & \textbf{$TD$} & \textbf{$DC_i$} & $v$(\%)\\
\hline
methods & 2876 & 11298 & .8165 & 81.7 & Rcpp & 1786 & 7333 & .5781 & 57.8 \\
utils & 2436 & 11197 & .8092 & 80.9 & magrittr & 930 & 5754 & .4536 & 45.4 \\
stats & 3953 & 10607 & .7666 & 76.7 & glue & 175 & 5271 & .4155 & 41.6 \\
grDevices & 1210 & 9302 & .6723 & 67.2 & digest & 215 & 5163 & .4070 & 40.7 \\
graphics & 1962 & 8923 & .6449 & 64.5 & R6 & 262 & 5075 & .4000 & 40.0 \\
Rcpp & 1786 & 7333 & .5301 & 53.0 & rlang & 612 & 5005 & .3946 & 39.5 \\
grid & 440 & 6930 & .5008 & 50.0 & pkgconfig & 7 & 4776 & .3765 & 37.7 \\
lattice & 400 & 6728 & .4861 & 48.6 & crayon & 172 & 4702 & .3707 & 37.1 \\
tools & 261 & 6113 & .4418 & 44.2 & assertthat & 205 & 4628 & .3648 & 36.5 \\
magrittr & 930 & 5754 & .4158 & 41.6 & stringi & 177 & 4623 & .3644 & 36.4 \\
Matrix & 874 & 5628 & .4072 & 81.6 & backports & 29 & 4599 & .3626 & 36.3 \\
glue & 175 & 5271 & .3810 & 38.1 & ellipsis & 12 & 4493 & .3543 & 35.4 \\
MASS & 1461 & 5242 & .3788 & 37.9 & vctrs & 14 & 4452 & .3514 & 35.1 \\
digest & 215 & 5163 & .3731 & 37.3 & zeallot & 6 & 4457 & .3514 & 35.1 \\
R6 & 262 & 5075 & .3667 & 36.7 & cli & 78 & 4436 & .3500 & 35.0 \\
rlang & 612 & 5005 & .3617 & 36.2 & fansi & 5 & 4437 & .3498 & 35.0 \\
pkgconfig & 7 & 4776 & .3451 & 34.5 & pillar & 19 & 4353 & .3431 & 34.3 \\
crayon & 172 & 4702 & .3398 & 34.0 & utf8 & 5 & 4357 & .3434 & 34.3 \\
assertthat & 205 & 4628 & .3344 & 33.4 & tibble & 673 & 4339 & .3420 & 34.2 \\
stringi & 177 & 4623 & .3340 & 33.4 & stringr & 841 & 4293 & .3384 & 33.8 \\
\hline
\end{tabular}
\end{table}

Packages with out-degree larger than 200 (so, packages that are transitively dependent on more than 200 other packages) can also be found in table~\ref{tab:4}. Again, this table shows how packages can show a limited number of imports but are indirectly (and probably, unknowingly) importing hundreds of transitive dependencies. The \textit{smartdata}~\citep{smartdata} package, for instance, might import a total of 24 packages (a large number already) but it ends up depending on 251 packages, ten times more than that, which means an inverse vulnerability (proportion of packages in the ecosystem that could break it) of 1.8\%. Extra care should be put in development of packages that either have high vulnerability or relatively large inverse one; developer efforts seem to reflect this tendency, as an statistically relevant relationship (p-value $\approx 0$) is found between the centrality of a package and the number of times it has been updated over time.

\begin{table}
\small
\centering
\caption{Packages with out-degree larger than 200 in CRAN.}
\label{tab:4}       
\begin{tabular}{|c|c|c|c|}
\hline
\multicolumn{4}{|c|}{\textbf{All packages (TC)}}\\
\hline
\textbf{Package} & Imports & Transitive & Inverse $v$(\%)\\
\hline
smartdata & 24 & 251 & 1.81 \\
ggstatsplot & 27 & 229 & 1.65 \\
psychNET & 23 & 206 & 1.49 \\
KNNShiny & 9 & 205 & 1.48 \\
STAT & 9 & 203 & 1.47 \\
CLUSTShiny & 8 & 201 & 1.45 \\
\hline
\end{tabular}
\end{table}

Controllability of complex networks~\citep{liu2011controllability} is another concept that could be useful to understand and characterize software package ecosystems. For scale-free networks with exponent $\gamma$ and average degree $\langle k \rangle$ we can use the following equation to compute the fraction of nodes that would be required to ``control" the system (known as driver nodes).

\begin{equation}
\frac{n_d}{n} \sim exp \left[-\frac{1}{2} \left(1-\frac{1}{\gamma-1}\right)\langle k \rangle \right]
\end{equation}

The results are shown in the last column of table~\ref{tab:2}. Note that, although the giant component network (GC) requires almost 79 packages to be controlled, it is misleading; once the transitive relationships are considered the number is notably lower. In summary, when the base packages are included, six packages would suffice to take control of the whole network; 16 in case these are not considered. This is consistent with the vulnerabilities found in table~\ref{tab:3}.

With regards to the second research question, thus, the present analysis of the packages available in CRAN using CNA quickly reveals how a few packages could expose the whole  ecosystem in case of failure, bug or malicious attack.

\subsection{RQ3: Modules}
\label{modules}

The links between packages in dependency networks are a product of the underlying patterns and structures. It could be expected, for instance, that packages that have similar functions are aggregated into package communities of relatively densely connected nodes. Finding a clear package community structure would mean that the software package ecosystem is highly modular (meaning that functions are basically independent from each other), while under-structured or disorganized projects would have a quasi-random structure. 

\begin{figure*}
\centering
  \includegraphics[width=\textwidth]{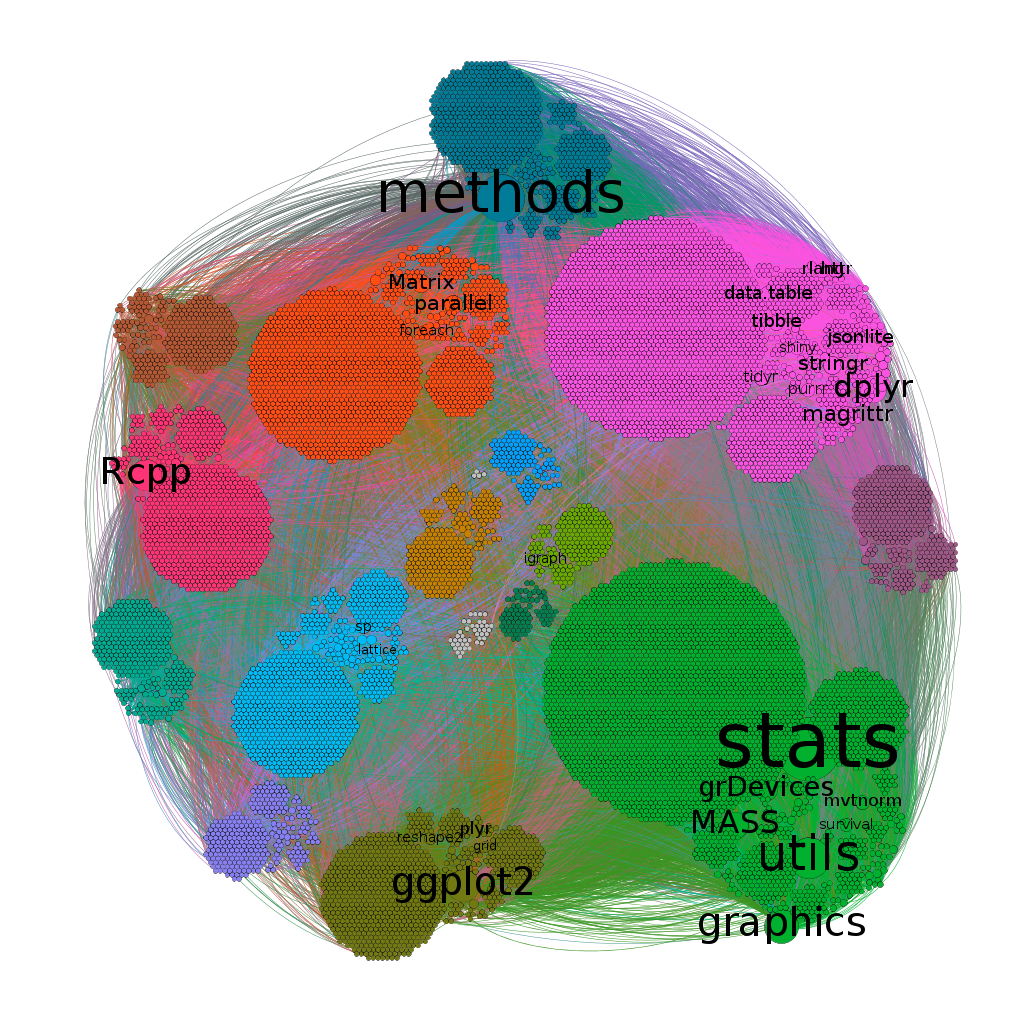}
\caption{Package communities obtained using Louvain's method for detecting community structure based on modularity. A total of 17 package communities are identified.}
\label{fig:comms}       
\end{figure*}

In general, one of the many challenges proposed in the field of complex network analysis consists of community detection, and multiple community detection algorithms have been described~\citep{gadek2018}. One of the most popular and widely used algorithm is the Louvain method~\citep{blondel2008fast}, which maximizes a modularity score for each community. For reference, the modularity of a partition is a scalar value between -1 and 1 that measures the density of links inside communities as compared to links between communities~\citep{newman2006modularity}. 

\begin{table}
\footnotesize
\centering
\caption{Summary of package communities found in the CRAN ecosystem, from largest to smallest (containing over 1\% of packages).}
\label{tab:5}       
\begin{tabular}{|p{0.05\textwidth}|p{0.15\textwidth}|p{0.15\textwidth}|p{0.19\textwidth}|p{0.35\textwidth}|}
\hline
\textbf{\%}&\textbf{Sample pkgs}&\textbf{Critical pkgs}&\textbf{Keywords in description}&\textbf{Summary}\\
\hline
\multirow{3}{*}{26.7} & mvtnorm & stats & Analysis & \multirow{3}{0.30\textwidth}{Popular packages for general statistical analysis.} \\
 & nlme & utils & Methods &  \\
 & lme4 & graphics & Regression &  \\
\hline
\multirow{3}{*}{19.9} & jsonlite & dplyr & API & \multirow{3}{0.30\textwidth}{Packages for managing and tidying data.}\\
 & tibble & magrittr & Tables &  \\
 & tidyr & stringr & Dataset &  \\
\hline
\multirow{3}{*}{11.9} & cluster & parallel & Selection & \multirow{3}{0.30\textwidth}{Classification, regression and clustering models.} \\
 & nnet & Matrix & Regression &  \\
 & caret & foreach & Classification &  \\
\hline
\multirow{3}{*}{6.94} & rgdal & sp & Spatial & \multirow{3}{0.30\textwidth}{Classes and methods for spatial data.}\\
 & fields & lattice & Raster &  \\
 & maptools & raster & Map &  \\
\hline
\multirow{3}{*}{6.68} & ggplot2 & ggplot2 & ggplot2 & \multirow{3}{0.30\textwidth}{Visualization tools and data arrangement.}\\
 & colorspace & grid & Plot &  \\
 & reshape2 & plyr & Tools &  \\
\hline
\multirow{3}{*}{6.35} & Rcpp & Rcpp & C++ & \multirow{3}{0.30\textwidth}{Integration languages into R, plus Bayesian and MCMC models.}\\
 & rstan & coda & Bayesian &  \\
 & rjags & Rdpack & MCMC &  \\
\hline
\multirow{3}{*}{5.90} & gsl & methods & Methods &  \multirow{3}{0.30\textwidth}{Tools allowing to analyze data with robust methods.}\\
 & rrcov & robustbase & Distribution &  \\
 & leaps & stats4 & Multivariate &  \\
\hline
\multirow{3}{*}{2.88} & rjson & XML & Text & \multirow{3}{0.30\textwidth}{Text mining applications plus web/java/json interfaces.} \\
 & tm & RCurl & Web &  \\
 & nlp & rjava & Java &  \\
\hline
\multirow{3}{*}{2.58} & expm & ape & Phylogenetic & \multirow{3}{0.30\textwidth}{Analysis of ecological/biological data in environmental sciences.}\\
 & ade4 & gtools & Species &  \\
 & seqinr & vegan & Trait &  \\
\hline
\multirow{3}{*}{2.55} & tseries & zoo & Time & \multirow{3}{0.30\textwidth}{Time series analysis and computational finance.}\\
 & timeData & xts & Series &  \\
 & timeSeries & forecast & Financial &  \\
\hline
\multirow{3}{*}{2.18} & Biobase & matrixStats & Gene & \multirow{3}{0.30\textwidth}{Bioconductor (bioinformatics) subcommunity.}\\
 & limma & R.utils & Genomic &  \\
 & Biostrings & future & RNA &  \\
\hline
\multirow{3}{*}{2.12} & tkrplot & rgl & GUI & \multirow{3}{0.30\textwidth}{Tools for both interactive GUI and (3D) graphics.}\\
 & tcltk2 & tcltk & Graphical &  \\
 & gWidgets & Rcmdr & 3D &  \\
\hline
\multirow{3}{*}{1.42} & sna & igraph & Network analysis & \multirow{3}{0.30\textwidth}{Tools for Social/Complex Network Analysis} \\
 & ergm & network & Graph &  \\
 & intergraph & GGally & Clustering &  \\
\hline
\end{tabular}
\end{table}

The Louvain algorithm is thus applied to the dependency network; the number of communities that emerge is stable at 17 and the resulting graph is shown in figure~\ref{fig:comms}, where the most relevant nodes (in regards to their in-degree) are labeled, obtaining a modularity of 0.4. The summary of the 13 communities that contain more than 1\% of the total packages can also be found in table~\ref{tab:5}, which reflects the portion of the total packages that each package community represents, three sample relevant packages (avoiding base packages that might distort the results) and three critical packages (understood as the ones with highest in-degree - highest vulnerabilities). 

To infer a meaning for each partition, natural language processing techniques were used; all the available textual descriptions for the packages in each set is aggregated and analyzed using \textit{spaCy} (https://spacy.io/), a Python library. After removing the common standard stopwords, the 30 top unique words found in the package's descriptions were annotated manually by three independent annotators, one with statistical and two with computer science background, that also analyzed each of the top packages in each package community, initially agreeing in 11 out of the 13 groups (84,6\%). The remaining two were discussed afterwards and a final agreement was reached with the identification found in Table~\ref{tab:5}, which produces a small summary of the structural reasons hidden behind the clustering produced by the algorithm.

Among them, the largest package community contains slightly over one quarter of the total packages in the ecosystem (26.7\%) and could be considered the functional core of the R package ecosystem, with the most popular packages for general statistical analysis. The rest of the communities are, overall, more specific of particular functionalities, disciplines or environments. Functionally, for instance, the second largest package community is formed by a large number of packages that are devoted to managing and cleaning data (such as the \textit{tidyverse} set) while the fifth largest group (6.68\%) is formed by ggplot2~\citep{ggplot2} and the visualization ecosystem (including all the ``gg" family) around this highly relevant package. In regards to disciplines, a few package communities are found that, for example, are focused in social/complex network analysis (1.42\%) or time series analysis (2.55\%). With the environmental perspective, there are at least two communities that are focused in environmental sciences, distinguished by whether their common packages are in the CRAN (2.58\%) or in the Bioconductor (2.18\%) repository.

Thus, and in response to the third research question, the modular analysis (using the community detection approach in the network of packages) reveals how this approach can detect and highlight the functional or environmental modules in a software package ecosystem such as CRAN.

\section{Discussion}
\label{discussion}

In this article, we empirically studied the CRAN software package ecosystem through complex network analysis tools, a method common in other fields but not as widely adopted for software engineering. Using the metadata from CRAN, the network of packages was built and its properties have been analyzed. 

\textbf{Structure and complexity}. Previous research already pointed out at the power-law (or near power-law) nature of dependency networks, although most of them was based in classes instead of packages~\citep{vsubelj2012}. In the case of CRAN, the degree distribution adheres to a power-law, both in degree and in in-degree, implying that packages with a high number of reverse dependencies tend to have a higher probability of receiving further incoming links, following the principles of the scale-free networks. Additionally, the small-world behaviour was tested for the CRAN case, noticing how the actual clustering of the network is orders of magnitude higher than the simulated clustering for a random network with the same characteristics. Average shortest patch between packages, on the other hand, is in the same order of the average shortest path in a random graph, which combined with the previous results in clustering determines that the behaviour of the CRAN package dependency network is in line with what is expected from a small-world network in software engineering: good relationship between packages that share similar functions while avoiding creating separate components far from each other. The small-world result for software package ecosystems has also been found, for instance, in the Debian repository~\citep{abate2009}. 

\textbf{Packages}. Our analysis of the packages available in CRAN (represented by the nodes of our network) reveals how a few packages could expose the whole ecosystem in case of failure, bug or malicious attack. When taking base packages into account, some of them (e.g. `methods', `utils' or `stats') could affect around 80\% of the packages in the repository if an update went wrong, as their transitive reverse dependencies are huge. However, it could be understood that base packages sit at the core of R and, therefore, the base packages and R could be considered as one. Additionally, it could be assumed that such base packages are both taken with more care and updated less often, so they might be less prone to a fault. In a deeper analysis, we found a total of 20 packages (besides the core 29 base packages) whose removal could impact more than 30\% of the other packages, which is a higher figure than found in other systems~\citep{vsubelj2012}. Some of them might seem harmless when only direct dependencies are considered, with less than 10 of them, but once transitive dependencies are considered, the potential impact could render the ecosystem unusable. In summary, we showed how CRAN has a few central packages (and a few highly imported ones) that are critical; high vulnerabilities, as shown in other ecosystems, should alert developers but, specially, maintainers, to keep a close look on potential bugs or security issues and their fixes for these packages, as being able to control a few of them would effectively give control over the ~\citep{kikas}. Moreover, from the opposite perspective, packages with large number of dependencies should also be monitored as their potential to fail is multiple times larger.

CRAN has strict policies on maintainers and contributions (https://cran.r-project.org/web/packages/policies.html). Among other policies, CRAN runs a periodic check on compatibility among packages; should any package fail the test, its maintainers would be notified and asked to resolve the issue before the following major R release, at the risk of having their package archived otherwise. CRAN also forces dependencies to be kept within itself or Bioconductor (to avoid external dependencies). Additionally, back-compatibility versions of current packages is not allowed, and any changes to CRAN packages that could cause significant disruption to other packages must be agreed with the CRAN maintainers before releasing it. These policies have a direct impact in mitigating most of the risks highlighted previously, although it is at the cost of the CRAN's maintainers efforts; developing tools based in CNA metrics could contribute towards minimizing CRAN volunteers lost time.

\textbf{Modules}. We also explored the structure of the CRAN package network from the perspective of community structure or modularity, running the Louvain algorithm and adding insights on the meaning of the resulting communities using NLP techniques on the available descriptions of packages. The main contribution here is to show how the dependency network obtained from CRAN reveals a significant package community structure and how such structure can be explained using the functionality or other relevant contextual aspects of the clustered packages. This is, thus, the proof of another property that is true for other networks as for software networks; hidden structure in software can be brought to surface using community detection algorithms~\citep{vsubelj2011}. In the case of CRAN, communities appear to be relatively balanced and, besides a core set of packages that cover the statistical analysis tools R is known for, there are multiple communities of packages devoted to common tasks (such as data wrangling) or to particular disciplines (such as bioinformatics). A modular approach has been shown to enhance functionality and evolvability~\citep{fortuna2011}. This results help drawing a map of CRAN, which, as a large software ecosystem, represents one of the most complex human made systems.

\section{Conclusion}
\label{conclusion}

Our analysis of the CRAN software package ecosystem from the perspective of complex network analysis shows how CRAN follows a scale-free and small-world behaviour, as found previously in other OSS package ecosystems, and that relates to good practices of software engineering. CRAN, however, presents a large number of packages that are critical for the correct function of the ecosystem and that, in case of any bug or issue, it could render the whole system unusable. CRAN's policies, in any case, are in place to prevent such event from happening, but it is something that has a maintenance cost. Finally, we also shown how the CRAN package network presents a significant modular structure, which is also a positive aspect of software engineering and enhances its functionality and evolvability, allowing such complex product of many contributors to go further into the future. Our findings show how package networks such as CRAN could benefit from complex network analysis as a tool to assess many aspects of software engineering, such as quality assurance or update risk evaluation. In particular, it is also worth highlighting how making dependency relationships more visible in package networks could help developers visualize the relevance of some packages and the critical value of others, so they could for example balance the number of dependencies or inverse dependencies, or aim to group together existing functionalities in order to reduce the number of critical packages to be maintained. In summary, CNA provides insights into relationships between components in package ecosystems that may be useful for a number of stakeholders, including core development teams, project managers and contributors (both individuals and organisations) who might want to understand how to contribute to the ecosystem in a way that best fits their audiences and interests.

\section{Threats to validity and future work}
\label{future}

The current work is limited, however, as package versions are not considered. CRAN's policies periodically checks compatibility among packages so only the packages need to be considered. The generalization of the results found previously would need to add package versioning (and, therefore, vulnerability per version) as an additional layer. In the same line, it would be interesting to extend the CNA analysis to study how the CRAN network has changed over time and how the maintainers' mitigation activities have developed over time in response to changes in the networks of dependencies. 

Another limitation is the use of manual analysis in section \ref{modules}, where NLP techniques were combined with manual annotation to infer a meaning for the clustering produced by the algorithm. Although the use of three independent reviewers mitigates the impact partially, it is acknowledged that experts from other fields (e.g. bioinformatics) might identify further nuances that were not captured in our analysis.

These findings show how CNA can be a valuable tool to study package ecosystems from the perspective of their dependency networks. Future work should follow at least three lines of research. First of all, the relationship between the CNA metrics of the packages and their quality metrics (e.g. open issues, user downloads, number of developers, binary sizes, among others) should be explored deeper. On one hand, to provide specific tools that developers could use to be aware of the most balanced approach for their packages, whether it is a leaner package with fewer dependencies or a more ``feature-full'' package directed to a particular audience, and to distinguish between those packages that are safe or those that should be avoided. On the other hand, the awareness of the developers should also be studied: what is their approach in regards to dependencies? Do they use any quantitative approaches when deciding on the dependencies they are going to include in their software? Second, the analysis could be extended with additional measures and network metrics, that could also be further combined with techniques from other disciplines such as NLP. Finally, CRAN is a popular ecosystem, but there are many more that could either be analyzed individually or compared from a complex network perspective to bring additional and potentially valuable findings to the field. 



\bibliographystyle{elsarticle-harv} 
\bibliography{bibliography}





\end{document}